\documentstyle[12pt]{article}

\begin{document}
\newcounter{foot}
\newcommand{\foot}{\addtocounter{foot}{1}\ $^{\!\!\thefoot}$}
\newcommand{\di}{\displaystyle}
\newcommand{\bm}{Brownian-motion}
\newcommand{\bp}{Brownian-particle}
\newcommand{\bps}{Brownian-particles}
\newcommand{\gf}{Gibbs-factorial}
\newcommand{\gfs}{Gibbs-fac\-to\-ri\-als}
 
\begin{center}
{\Large\bf On the Interrelation between}\\
{\Large\bf Gibbs Hypothesis}\\
{\Large\bf and Symmetry Postulate}\\
P.\ Hrask\'o\protect\footnote{E-mail:hraskop@ceu.hu}\\
{\em Department of Theoretical Physics, Janus Pannonius
University\\
P\'ecs, Ifjus\'ag u. 6, 7604 P\'ecs, Hungary}
\end{center}
 
\vspace{8mm}
 
{\bf Abstract:} The question of the origin of the \gfs\ is
reconsidered. It is argued, using the example of the \bm , that
their appearance in classical statistical calculations cannot
in general be traced back to the symmetry postulate of quantum
mechanics as often assumed.
 
 
\vspace{8mm}
 
In this note I propose to consider a problem of interpretation
with no immediate practical consequence. It seems, nevertheless,
worth of clarification.
 
In classical mechanics of pointlike particles dynamical states
are associated with points of the phase-space.
Though in general different points in phase-space correspond to
different dynamical states of the system, Gibbs proposed\foot\ to
make an important exception to this rule: phase points which
differ only by {\em permutations} of identical particles
are to be taken to
represent the same dynamical state and, therefore, must be counted
only once in probability distributions over
phase-space.
At the same time Gibbs suggests that since integrals over these
distributions are much
easier to handle without the above restriction, an equivalent but
simpler procedure would consist in dividing the unconstrained
integrals
by the number of equivalent phase-space points. This last quantity
is easy to compute: it is given by a simple product each term of
which is equal to the factorial of the number of particles of a
given type. The factorials in this product are the
{\em\gfs } and the whole procedure will be called below {\em Gibbs
hypothesis}.
 
The necessity of this strange statistics
was recognized by Gibbs in connection with the mixing of
gases and liquids. It follows from thermodynamics that mixing
of different substances is accompanied by an entropy increase
related to the work extractable from the mixing process.
However, the statistical description
of the process reproduces this increase of the entropy only if
\gfs\
are properly taken into account; otherwise the theory would make
no distinction
between mixing of different or identical gases and would give no
entropy change in either case.
This argument of Gibbs may also be expressed in a somewhat
more general form by saying that methods of
statistical thermodynamics lead to an additive (extensive)
entropy only if \gfs\ are introduced\foot .
 
Though the necessity of using \gfs\ in calculations had been
quickly accepted, sharp controversy as to the deeper sense of this
procedure arose in the first quarter of the
century\foot .
However, with the birth of modern quantum mechanics this debate
came to a sudden end, since the {\em symmetry postulate} of
quantum theory seemed to settle
the problem once for all. This fundamental principle requires
that under
permutation of identical particles the state vector changes at
most by a phase factor which leads to no change at all in the
physical state of the system. Therefore, whenever symmetry
postulate is applicable it provides firm and transparent basis for
Gibbs' procedure indeed.
This conclusion is corroborated by the
theories of ideal Bose and Fermi gases which in high temperature
limit go over to the classical statistical description of ideal
gases with the \gf\ automatically included\foot .
 
According to the general opinion, the symmetry postulate of
quantum mechanics fully elucidates the origin of Gibbs hypothesis
and, therefore, serves as the  {\em general explanation} of it.
In order to illustrate this point of view I quote two examples.
The first one is from the Part 6.6 of\foot :
"It is not possible to understand classically why we must
divide $\sum (E)$ by $N!$ to obtain the correct counting of
states. The reason is inherently quantum mechanical..." The second
is from the Chapter 16 of the standard text by T.L.Terrel referred
to above: "The new factor $1/N!\cdots$ has arisen from
our consideration of the symmetry of wave functions. In turn, the
quantum mechanical postulate on symmetry of wave functions has its
origin in the experimental indistinguishability of identical
particles. Gibbs had the intuitive forsight, before the advent of
quantum mechanics, to insert this factor $1/N!$."
 
A closer reflection, however, reveals that derivability of the
\gfs\ from the symmetry postulate is in itself insufficient to
provide an explanation of Gibbs hypothesis in the general case.
It certainly
ensures that quantum mechanics of {\em identical particles}
automatically obey this hypothesis but tells nothing of systems,
consisting of {\em similar} (i.e. practically indistinguishable
but not truly identical) {\em particles} as e.g. colloid particles
in a
suspension. Systems, consisting of such constituents, fall outside
the scope of the symmetry postulate but, nevertheless, require
\gfs\
for their correct description in the framework of statistical
thermodynamics.
 
\gfs\ enter into the working formulae of colloid physics somewhat
implicitely. They are hidden within the relations taken over from
the theory of dilute molecular solution whose properties are
closely related to those of dilute suspensions. In order to
expound the significance of \gfs\ for colloids as
systems, consisting of similar but certainly not identical
particles, let us consider
the widely known example of the equilibrium of \bps\ in the
gravitational field of the Earth which was the object of
J.Perrin's famous experiments. The treatment of this phenomenon in
colloid physics\foot\ is based on the formula
\begin{equation}
RT\,d\ln a' + gv'(\rho '-\rho )\,dh = 0,\label{6a}
\end{equation}
in which $a',\;v'$ and $\rho '$ are the
activity, molar volume and weight-volume density of the dissolved
substance, $\rho$ is the density of the solvent and $h$ is the
height above some reference plane.
For ideal solutions of constant density this equation may be
integrated to obtain the barometric formula for the particle
density $C'$
\begin{equation}
C' = C'_0\exp{\left [ -{gv'(\rho '-\rho )\over RT}\cdot
h\right ]},\label{7a} \end{equation}
which was used by Perrin to calculate the Avogadro's number.
 
Where does formula (\ref{6a}) come from? It comes from
the formula for the chemical potential of the solute
in
a {\em molecular} solution applied to colloid particles in a
suspension. Such a borrowing conforms with the basic
hypothesis of Einstein's famous 1905 paper\foot\ to the effect
that
\bps\ are nothing but molecules of visible size.
 
The second term in
(\ref{6a}) is the change in the gravitational potential energy of
the dissolved
matter corrected for the hydrostatic lifting force. What concerns
the logarithmic term, in textbooks on physical chemistry\foot\
it is
related through equilibrium condition $\mu '(l) = \mu '(g)$
to the logarithmic term in the chemical potential of an
ideal gas.
 
Though these considerations constitute a true explanation of
(\ref{6a}), an alternative
approach\foot\ reveals that the
logarithmic term bears in fact its origin from the
\gf\ $n'!$ of the solute particles.
The starting point of this latter approach is the general
expression of the free enthalpy $G$ of a dilute solution (in the
limit $n'\longrightarrow 0$) which obeys all the basic physical
requirements known to be satisfied by free enthalpy.
 
If at $n'=0$ $G$ were a smooth function
of $n'$ it could be expanded in powers of $n'$ and to lowest
nontrivial order the simple expression
\begin{equation}
G = G^* + n'\cdot\alpha\label{8a}
\end{equation}
would be obtained in which $G^*$ is the free enthalpy of
the pure solvent and $\alpha$ is independent of $n'$. But
$G$ is in fact a singular function of $n'$ at
$n'=0$ and cannot be expanded in a Taylor-series around this
point.
The origin of the singularity is the \gf\ $n'!$ in the denominator
of the statistical sum $Z$. Since $G=-kT\,\ln Z$ the free enthalpy
does contain a term $+kT\cdot\ln n'!$ which can be written also
in
the form $kT\cdot n'(\ln n'-1)$. It is this term which cannot be
expanded around $n'=0$ and so must be explicitely included into
(\ref{8a}). Hence, we have at $n'\longrightarrow 0$ the expression
\begin{equation}
G = G^* + n'\cdot\alpha + kT\cdot n'(\ln n'-1).\label{8b}
\end{equation}
 
However, this formula is still in flaw. $G$ must be a first
order homogeneous function of $n$ and $n'$, the number of
molecules in the solvent and the solute respectively. Since the
term $\ln n'$ spoils homogeneity, $G$ necessarily contains another
logarithmic term too wich combine with $\ln n'$ into $\ln (n'/n)$.
The general lowest order form of the free enthalpy is, therefore,
\begin{equation}
G = G^* + n'\cdot\alpha + kT\cdot n'\left [\ln
\left ({n'\over n}\right )-1\right ].\label{9a}
\end{equation}
 
From this formula one can calculate the chemical potentials per
mole of both the solvent and the solute:
\begin{eqnarray*}
\mu =&&\hspace{-6mm} N_A{\partial G\over\partial n} =
\mu^*-RT{n'\over n}\\
\mu' =&&\hspace{-6mm} N_A{\partial G\over\partial n'} =
N_A\alpha + RT\,\ln\left ({n'\over n}\right ).
\end{eqnarray*}
In calculating $\mu$ we put $\di {\partial\alpha\over\partial
n}=0$ since, owing to homogeneity, the independence of $\alpha$ on
$n'$ implies its independence on $n$ too.
 
In the gravitational field of the Earth these equations are
modified to
\begin{eqnarray}
\mu =&&\hspace{-6mm}
\mu^*-RT{n'\over n}+gv\rho h\label{9d}\\
\mu' =&&\hspace{-6mm}
N_A\alpha + RT\,\ln\left ({n'\over n}\right )+gv'\rho
'h.\label{9e}
\end{eqnarray}
The constancy of the chemical potentials along the vertical
direction leads to the equations
\begin{eqnarray}
{\partial\mu^*\over\partial P}{dP\over dh} - RT{d\over dh}\left
({n'\over n}\right ) + gv\rho =&&\hspace{-6mm} 0\label{9b}\\
N_A{\partial\alpha\over\partial P}{dP\over dh} + RT{d\over
dh}\ln\left ({n'\over n}\right ) + gv'\rho ' =&&\hspace{-6mm}
0.\label{9c}
\end{eqnarray}
For the pure solvent $\mu^* = N_AG^*/n$, hence
\[{\partial\mu^*\over\partial P} = {N_A\over n}{\partial
G^*\over\partial P} = {N_A\over n}V = v.\]
On the other hand, from (\ref{9a}) we have
\[N_A{\partial\alpha\over\partial P} = {N_A\over n'}\left
({\partial G\over\partial P}-{\partial G^*\over\partial P}\right )
= {N_A\over n'}V' = v'.\]
Substituting these expressions into (\ref{9b}) and (\ref{9c}) we
have (after replacing $n'/n$ with the ratio of the particle
densities)
\begin{eqnarray}
v{dP\over dh} - RT{d\over dh}\left ({C'\over C}\right ) + gv\rho
=&&\hspace{-6mm} 0\label{10a}\\
v'{dP\over dh} + RT{d\over dh}\ln\left ({C'\over C}\right ) +
gv'\rho ' =&&\hspace{-6mm}0.\label{10b}
\end{eqnarray}
Now, $C$ is practically independent of $h$, hence
\[{d\over dh}\left ({C'\over C}\right )= {1\over C}{dC'\over
dh}\hspace{5mm}\mbox{and}\hspace{5mm}
{d\over dh}\ln\left ({C'\over C}\right )= {1\over C'}{dC'\over
dh}.\]
The first of these expressions is $C'/C$-times smaller than the
second and can be neglected. Then from (\ref{10a}) we have
$\di{dP\over dh}=-g\rho$. Substituting this into (\ref{10b}) we
obtain the equation
\begin{equation}
RT{d\over dh}\ln\left ({C'\over C}\right )+g(\rho '-\rho
)v'=0,\label{last}
\end{equation}
which when solved leads again to the barometric
formula (\ref{7a}).
 
In the second approach which has been now completed it is formula
(\ref{last}) which corresponds to (\ref{6a}). Since in the limit
$n'\longrightarrow 0$ this last formula becomes identical to
(\ref{last}) it is
plainly obvious that the logarithmic terms
correspond to each other and both originate from the last term
of (\ref{8b}) which in turn stems from the \gf\ $n'!$. If,
therefore, one
admits that the application of formulae of molecular physics
to suspensions --- the common practice in colloid physics ---
is justified, then one thereby accepts also that \gfs\ must be
applied to systems of both identical and similar particles. Since
the latters are outside the domain of the symmetry postulate,
quantum mechanics can not be claimed to provide the general
basis for the explanation of Gibbs hypothesis
in spite of the common belief.
 
The notion of similar particles involves obviously an element of
subjectivity not shared by the concept of identity of particles.
However, Gibbs-factorials appear always as ingredients of the
entropy (or related thermodynamic functions), and the ambiguity in
the distinction between similar particles and distinguishable
ones leads to the same kind of uncertainty in it
that has already
been thoroughly discussed earlier in connection with the
anthropomorphic notion of the entropy\foot .
 
This latter concept is based on the information theoretic
interpretation which identifies entropy with the measure of
the lack of information about the system\foot . But information
can be quantified only with respect to some set of a priori
expectations and it is
these expectations through which an unavoidable
anthropomorphic element comes into play.
A good example is provided by the existence of isotopes\foot .
So far as the differences in their properties are beyond the
capacity of experiments (as is the case in classical chemistry) no
inconsistency is introduced by calculating the entropy of elements
as if isotopes were
indistinguishable, though from the point of view of the theory of
isotope separation this would be a crude error.
 
The author is indebted to Dr.R.Schiller for stimulating
discussions on the subject.
 
\vspace{5mm}
 
{\large\bf Notes}
\setcounter{foot}{0}
 
\foot\ J.W.Gibbs, {\em Elementary Principles in Statistical
Mechanics,} (Yale University Press, New Haven, 1902), Chapter 15.
 
\sloppy
\foot\ C.Kittel, {\em Elementary Statistical Physics,} (New York,
John Wiley\& Sons), Chapter 9.
 
\sloppy
\foot\ Interesting details on the history of the \gf\ are found in
M.J.Klein, "Ehrensfest's contributions to the Development of
Quantum Statistics I. II.," Proc. Koninkl. Ned. akad. Wetensch.
(Amsterdam), {\bf 62,} 41 (1958).
 
\sloppy
\foot\ T.L.Terrel, {\em Statistical Mechanics,} (New
York, McGraw-Hill, 1956).
 
\foot\ K.Huang, {\em
Statistical Mechanics,} (New York, John Wiley\& Sons, 1987)
 
\foot\ G.Scatchard, {\em Equilibrium in
Solutions. Surface and Colloid Chemistry,} (Harvard University
Press, Cambridge, 1976) p. 256.
 
\foot\ A.Einstein, {\em Investigations on the Theory of Brownian
Movement,} (New-York, Dover, 1956)
 
\foot\ P.W.Atkins, {\em Physical Chemistry,} (Oxford University
Press, Oxford, 1990)
 
\foot\ L.D.Landau and E.M.Lifshitz, {\em Statistical Mechanics,}
(Pergamon Press, London, 1981), Chapters 87 - 90.
 
\foot The anthropomorphic nature of the entropy has
been stressed
by E.T.Jaynes, "Gibbs vs Boltzmann Entropies," Am. J. Phys. {\bf
33,} 391 (1965). The opposite point of view is defended in
K.G.Denbigh
and J.S.Denbigh, {\em Entropy in Relation to Incomplete
Knowledge,} (Cambridge University Press, 1985). For a critical
review of this book see H.Price, Brit. J. Phil. Sci. {\bf 42,}
111 (1991).
 
\foot L.Brillouin, {\em Science and Information
Theory,} (Academic Press, New York, 1956) Part 12.
 
\foot E.Schroedinger, "Isotopie und das Gibbs'sche
Paradoxon," Zeitschrift f\"ur Physik {\bf 5,} 163 (1921).

\end{document}